\documentclass[12pt]{article}
\usepackage{euscript}
\usepackage{amssymb}
\usepackage{amsmath}
\usepackage{graphicx}
\usepackage[small,sc,center]{titlesec}
\usepackage{indentfirst}

\newcommand{\nc}{\newcommand}
\nc{\ct}{\EuScript{C}_\mathrm{t}}
\nc{\dEt}{\dot{\EuScript{E}}_\mathrm{t}}
\nc{\dUt}{\dot{\EuScript{U}}_\mathrm{t}}
\nc{\dr}{\EuScript{D}_\mathrm{rad}}
\nc{\dt}{\EuScript{D}_\mathrm{t}}
\nc{\ek}{E_\mathrm{K}}
\nc{\et}{E_\mathrm{t}}
\nc{\Fc}{\mathbf{F}_\mathrm{c}}
\nc{\Fr}{\mathbf{F}_\mathrm{r}}
\nc{\Ft}{\mathbf{F}_\mathrm{t}}
\nc{\hp}{H_\mathrm{P}}
\nc{\K}{\,\mathrm{K}}
\nc{\mbol}{M_\mathrm{bol}}
\nc{\mzams}{M_\mathrm{ZAMS}}
\nc{\pd}[2]{\dfrac{\partial{#1}}{\partial{#2}}}
\nc{\pt}{P_\mathrm{t}}
\nc{\st}{\EuScript{S}_\mathrm{t}}
\nc{\Teff}{T_\mathrm{eff}}
\nc{\xhec}{X({}^4\mathrm{He})_\mathrm{c}}
\begin{document}

\begin{center}
\textbf{EVOLUTION AND THE PERIOD--LUMINOSITY RELATION FOR RED SUPERGIANTS IN THE MAGELLANIC CLOUDS}

\textbf{Yu. A. Fadeyev\footnote{E--mail: fadeyev@inasan.ru}}

\textit{Institute of Astronomy, Russian Academy of Sciences, Pyatnitskaya ul. 48, Moscow, 109017 Russia}

Received December 5, 2012

\end{center}

\textbf{Abstract} ---
Excitation of radial pulsations in red supergiants of Magellanic Clouds is investigated
using the stellar evolution calculations and the self--consistent solution of the equations
of radiation hydrodynamics and turbulent convection.
The stars with initial masses $6M_\odot\le\mzams\le 28M_\odot$ and the initial chemical
composition $X=0.7$, $0.004\le Z\le 0.008$ are shown to be unstable against fundamental
mode oscillations with periods from 17 to 1200 days as they become helium burning
red supergiants.
The period--luminosity relation slightly depends on the mass loss rate varying
with a factor of three, whereas its dependence on the metal abundance is given by
$\delta\mbol = 0.89 \delta\log Z$.
In comparison with galactic red supergiants the low metal abundances in red supergiants
of Magellanic Clouds are responsible for their higher effective temperatures and
substantially narrower ranges of evolutionary radius change during helium burning .
Therefore on the period--mass diagram the red supergiants of Magellanic Clouds are
located within the strip with width of $\delta\log M\approx 0.09$, so that
the uncertainty of mass evaluation of the red supergiant with the known
pulsation period is nearly 25\%.


\section*{introduction}

Red supergiants belong to brightest stars ($L\lesssim 3\times 10^5L_\odot$)
and are also known as long--period variables (LPV) with cyclic light variations
on the time scale from a few dozen to $\sim 10^3$ days.
The pulsational nature of light variability in red supergiants has been firstly
shown by Stothers (1969) and later was supported in several following works
(Stothers 1972; Li and Gong 1994; Heger et al. 1997).
In the General Catalogue of Variable Stars (Samus et al. 2011) these stars are
classified as pulsating variables of type SR with semiregular light variations.

The pulsation hypothesis for the light variability of red supergiants is strongly
supported by the period--luminosity relation derived by Glass (1979) from
photometric observations of 38 brightest M--type supergiants in the Large and Small
Magellanic Clouds (LMC and SMC).
Since then the number of semiregular late--type variables observed in these stellar
systems increased by several times (Yang and Jiang 2011, 2012).
The period--luminosity relation for red supergiants is of great importance due to the
fact that in comparison with Cepheids they are more appropriate for extragalactic distance
calibration (Pierce et al. 2000; Jurcevic et al. 2000).

However in contrast to Cepheids with well known mechanism of pulsation instability
the red supergiants still remain scarcely studied pulsating variable stars.
Theoretical studies of pulsational instability in red supergiants were presented
only in two papers by Li and Gong (1994) and Guo and Li (2002) 
devoted to the linear nonadiabatic analysis of radial oscillations.
The authors concluded that the observed variability of red supergiants is due to
radial oscillations in either the fundamental mode or the first overtone
and pulsations are driven by the $\kappa$--mechanism in the helium ionizing zones.
Unfortunately, the linear analysis employed in these studies is based on the simplified
treatment of convection and does not take into account effects of turbulence
(Li 1992a, 1992b).
In red supergiants the turbulence plays an important role because of the
large extension of the outer convective zone (as large as 70\% of the stellar mass)
as well as due to substantial contribution of the turbulent pressure which is as large as one third
of the total thermodynamic pressure (i.e. the sum of the gaseous and radiation pressure).

Hydrodynamic calculations of nonlinear radial oscillations in red supergiants based on
the self--consistent solution of the equations of radiation hydrodynamics and turbulent
convection were done by Fadeyev (2012).
This study deals with stars with initial masses $8M_\odot\le\mzams\le 20M_\odot$ and
relative mass abundances of hydrogen and elements heavier than helium (metals)
$X=0.7$ and $Z=0.02$, respectively.
However, this composition is typical for Population~I stars of our Galaxy, whereas
in stars of LMC and SMC
the metal abundancies are significantly lower: $0.003\le Z\le 0.008$
(Peimbert and Torres--Peimbert 1974, 1976).
In the present paper we show that lower metal abundances in red supergiants of LMC and SMC
lead to effective temperatures higher by $300\K\le\Delta\Teff\le 500\K$
in comparison with galactic red supergiants.
In late--type stars ($\Teff\approx 3500\K$) such a difference in
effective temperatures leads to significant changes in both stability of the star against
radial oscillations and the pulsation period.

The goal of the present work is to study the pulsational properties of red supergiants in
LMC and SMC with methods of hydrodynamic computations of nonlinear radial stellar
oscillations.
We consider the stars with initial masses $6M_\odot\le\mzams\le 28M_\odot$
and relative mass abundances of hydrogen and metals on the zero age main sequence
(ZAMS) $X=0.7$ and $Z=0.004$, 0.008, respectively.

\section*{the method of computations}

\subsection*{stellar evolution}

The study of self--exciting stellar oscillations is the Cauchy problem
for equations of radiation hydrodynamics and turbulent convection that describe
the spherically--symmetric motion of the self--gravitating gas.
Initial conditions in such a problem are the spatial distributions of physical variables
corresponding to the hydrostatic equilibrium.
In the present work the initial conditions are obtained from the stellar evolution
calculations.
Description of the methods employed are given in our previous papers
(Fadeyev 2007, 2010).

In calculations of stellar evolution we used the steady--state local convection model
(B\"ohm--Vitense 1958) with the ratio of mixing length to pressure scale height
$\ell/\hp = 1.6$.
It was also assumed that the size of the convective core increases due to convective
overshooting by $0.1\hp$.

Stars with initial masses $6M_\odot\le\mzams < 15M_\odot$ occupy the red supergiant
domain only during the initial phase of core helium burning and later they leave
this domain as helium is exhausted in the convective core.
For example, the star $\mzams=10M_\odot$, $Z=0.008$ leaves the red supergiant
domain when the central helium abundance is $X({}^4\mathrm{He})_\mathrm{c} < 0.4$.
In such a case the evolutionary track loops the Hertzsprung--Russel (HR) diagram
and crosses the Cepheid instability strip.
More massive stars ($15M_\odot\le\mzams < 28M_\odot$) remain red supergiants during
the whole stage of thermonuclear core helium burning and the shape of their
evolutionary tracks in the HR diagram depends on the mass loss rate $\dot M$.

Various empirical formulae for $\dot M$ were recently discussed by Mauron and Josselin (2011)
who concluded that the best agreement with observations of red supergiants in
Magellanic Clouds is obtained with the formula by de Jager et al. (1988).
In the present study the mass loss rate of evolving red supergiants was assumed to be
$\dot M = f \dot M_\mathrm{NJ}$, where $\dot M_\mathrm{NJ}$ is the mass loss rate
evaluated by formula of Nieuwenhuijzen and de Jager (1998) which only
insignificantly differs from that by de Jager al. (1988).
The factor $f$ is introduced to estimate the dependence of results of
hydrodynamic computations on uncertainties in mass loss rates and
in stellar evolution calculations was fixed in the range $0.3\le f\le 1$.

The role of the mass loss rate in the evolutionary track of the red supergiant with
initial mass $\mzams = 20M_\odot$ and metal abundances $Z=0.004$, 0.008 and 0.02
is illustrated in Fig. 1.
Of most interest are parts of the track that correspond to the helium core burning
and in Fig. 1 they are shown by solid lines.
Dottedd lines correspond to the evolutionary stage when the energy source in
the stellar center is gravitational contraction of the core
and movement in the HR diagram is several orders of magnitude faster than during
thermonuclear helium burning.
We do not discuss the stage of gravitational core contraction because of
negligible probability to observe such stars.

\subsection*{nonlinear stellar pulsations}

In solution of the equations of hydrodynamics we assumed that the flux
of enthalpy $\Fc$ and the flux of the turbulent energy of convective
elements $\Ft$ are defined according to Kuhfu\ss (1986).
In contrast to previous works of the author (Fadeyev 2011, 2012)
the equation of motion and the energy equation are written in the
following form:
\begin{gather}
\label{eqmot}
\pd{{}^2r}{t^2} = - \dfrac{G M_r}{r^2} - 4\pi r^2 \pd{}{M_r}\left(P + \pt\right) + \dUt ,
\\[5pt]
\label{enereq1}
\pd{E}{t} + P\pd{V}{t} = - \dfrac{1}{\rho} \nabla\cdot \left(\Fr + \Fc\right) - \EuScript{C}_\mathrm{t} ,
\\[5pt]
\label{enereq2}
\pd{\et}{t} + \pt \pd{V}{t} = - \dfrac{1}{\rho} \nabla\cdot \Ft + \dEt + \EuScript{C}_\mathrm{t} .
\end{gather}

The equation of motion (\ref{eqmot}) describes the change of the radius $r$
and velocity $U=\partial r/\partial t$ of the mass zone with Lagrangean coordinate
$M_r$ due to gravity (here $G$ is the gravitational constant) and gradients of the
total thermodynamic pressure $P$ (the gas pressure and radiation pressure)
and the turbulent pressure $P_\mathrm{t}$.
The last term in the right--hand--side of equation (\ref{eqmot}) accounts for the
momentum transfer between the gas flow and turbulent elements.

Equations (\ref{enereq1}) and (\ref{enereq2}) describe the change of the specific internal
energy of gas $E$ (the sum of the translational energy of gas particles, excitation and
ionization energy of atoms, energy of radiation) and the specific turbulent kinetic
energy $\et$.
Here $\rho = 1/V$ is the gas density, $V$ is the specific volume, $\Fr$ is the
radiation flux calculated in approximation of radiation heat conduction.
The coupling term $\ct$ is defined as
\begin{equation}
\ct = \st - \dt - \dr ,
\end{equation}
where $\st$ is the rate of specific turbulent energy generation due to buoyancy forces,
$\dt$ and $\dr$ are the rates of turbulent energy dissipation due to molecular
viscosity and radiation, respectively.
Formulae for these quantities are given by Wuchterl and Feuchtinger (1998).

The turbulence is assumed to be isotropic, so that viscous stresses are determined by
the Reynolds stress tensor (see, e.g., Pope (2000)).
In spherical geometry the rates of momentum transfer $\dUt$ and energy transfer
$\dEt$ between gas flows and turbulent elements are scalar trace--free parts
of the Reynolds tensor (Wuchterl and Feuchtinger 1998).

\section*{results of hydrodynamic calculations}

\subsection*{pulsational instability domain}

Some models of evolutionary sequences corresponding to the thermonuclear core helium
burning were used as initial conditions in the Cauchy problem for equations
(\ref{eqmot}) -- (\ref{enereq2}).
The role of small perturbations was played by errors arising in interpolation
of the evolutionary model computed by the Henyey method to the hydrodynamical
model with Lagrangean mass intervals increasing inward as geometric progression.
To diminish interpolation errors the evolutionary calculations were done with
the number of mass zones ranging from $5\times 10^3$ to $10^4$, whereas in
hydrodynamical models the number of Lagrangean intervals was $N\approx 10^3$.

Integration of the equations of hydrodynamics (\ref{eqmot}) -- (\ref{enereq2})
with respect to time $t$ was accompanied by evaluation of the kinetic energy
\begin{equation}
\ek(t) = \frac{1}{2}\sum\limits_{j=2}^N \Delta M_{j-1/2} U_j(t)^2 ,
\end{equation}
where $U_j$ is the gas flow velocity in the $j$--th Lagrangean zone,
$\Delta M_{j-1/2}$ is the mass interval between ($j-1$)--th and $j$--th
zones.
The value $j=1$ corresponds to the inner boundary of the hydrodynamical
model where $U_1=0$ and the luminosity satifies the condition
$\partial L_1/\partial t = 0$.

Exponential decrease of the averaged over cycle kinetic energy $\ek(t)$
shows that the star is stable against radial pulsations.
For the criterion of pulsational stability we used the condition that
the averaged over cycle kinetic energy decreases by at least two orders of
magnitude in comparison with kinetic energy of the initial perturbation.

In the case of the growth of kinetic energy the hydrodynamic calculations
were carried out untill the limit cycle is attained and the averaged
over cycle kinetic energy $\ek$ becomes independent of time.
This condition is fulfilled for models with initial masses $\mzams\le 15M_\odot$
due to small nonlinear effects.
In more massive stars the amplitude of limit cycle oscillations varies
from cycle to cycle and the condition of the constant cycle--averaged kinetic energy
is fulfilled only approximately within sufficiently large time intervals.

The mean period of radial pulsations $\Pi$ was evaluated from the discrete
Fourier transform of the temporal dependence $\ek(t)$ for both pulsationally
stable and pulsationally unstable models.
Within the initial time interval with linear change of $\ln{\ek}_{\max}$
the instability growth rate $\eta = \Pi d\ln{\ek}_{\max}/dt$ was evaluated,
where ${\ek}_{\max}$ is the maximum kinetic energy attained within one
oscillation cycle.
For all hydrodynamical models with positive growth rate ($\eta > 0$)
stellar pulsations were found to be in the fundamental mode.

The change of the instability against radial oscillations in the evolving red supergiant
is illustrated by plots of $\eta$ in Fig. 2 for several values of $\mzams$.
The plots are shown versus the relative mass abundance of helium in the
stellar center $\xhec$ because duration of thermonuclear helium burning depends
on the stellar mass.
In all models the onset of the core helium burning takes place for $0.97\le\xhec\le 0.99$.

The growth rate of pulsational instability depends on many factors
but the most important is the size of the outer convective zone.
Luminosity decrease of red supergiants $15M_\odot < \mzams\le 20M_\odot$
during the initial helium burning (see Fig. 1) is accompanied by
shallowing of the outer convective zone.
For $\xhec\approx 0.4$ the mass of the convective zone reaches its minimum
(nearly a half of the stellar mass) and as is seen from plots in Fig.~2 for
$\mzams = 18M_\odot$ and $\mzams = 20M_\odot$
the pulsational instability growth rate becomes highest.
Further helium exhaustion is accompanied by increase of the both luminosity
of the red supergiant and mass of the outer convection zone, whereas
the instability growth rate $\eta$ decreases.

Typical plots of the gas flow velocity at the upper boundary $U$
and the bolometric light $\mbol$ of the red supergiant with nearly
maximum growth rate $\eta$ are shown in Fig. 3.
The initial stellar mass, the factor of the mass loss rate and the central
helium abundance are $\mzams = 20M_\odot$, $f=0.3$ and $\xhec=0.39$,
respectively.
For the sake of comparison the plots of $U$ and $\mbol$ for the model
$\mzams = 9M_\odot$, $\xhec=0.36$ are shown in Fig. 4.

Instability against radial oscillations enhances with increasing $\mzams$
due to higher stellar luminosity and increasing nonadiabaticity of
stellar pulsations.
This is illustrated in Fig. 2 by the plot for the model sequence $\mzams=25M_\odot$.
Here however one should be noted that for $\eta > 0.2$ estimates of this
quantity become incorrect because of the small number of maxima of the
kinetic energy during the stage of instability growth.

\subsection*{period--luminosity relation}

The change of the pulsation period $\Pi$ of the evolving red supergiant is mainly due
to evolutionary changes of its radius $R$ because the period of the fundamental mode
is proportional to the sound travel time from the stellar center to the surface.
In the beginning of thermonuclear helium burning the luminosity of the red
supergiant decreases and the effective temperature increases (see Fig.~1)
therefore the fundamental mode period diminishes.
Following increase of luminosity and decrease of effective temperature are
accompanied by increase of the period $\Pi$.
Such a general property of red supergiants is illustrated in Fig.~5 for three models
with initial mass $\mzams=20M_\odot$ and three different metal abundances $Z$.
Each evolutionary sequence of hydrodynamical models is represented by filled circles
connected by solid lines for the mass loss rate factor $f=0.3$ and by dotted lines
for $f=1$.

The period--luminosity diagram in Fig.~5 reveals remarkable differences between red
supergiants of our Galaxy and those of Magellanic Clouds.
First, for the fixed luminosity galactic red supergiants pulsate with longer periods.
As noted above this is due to the fact that red supergiants with lower metal abundances
have higher effective temperatures and smaller radii.
Second, ranges of evolutionary changes of radial oscillation periods of galactic red
supergiants are appreciably wider.
For example, for stars with initial mass $\mzams=20M_\odot$ and $Z=0.02$
the pulsation period changes in the range from 510 to 1020 days.
With decreasing $Z$ the range of evolutionary changes of the radial oscillation
period shrinks to 20\% and is
$380~\mathrm{day}\le \Pi\le 460~\mathrm{day}$ for $Z=0.008$
and $370~\mathrm{day}\le\Pi\le 430~\mathrm{day}$ for $Z=0.004$.
The cause of such remarkable differences between red supergiants of our Galaxy and
those of Magellanic Clouds is due to the ranges of evolutionary changes of
the stellar radius.
For example, in galactic red supergiants with $\mzams=20M_\odot$ the evolutionary
change of the radius $R$ is about one third, whereas in red supergiants of Magellanic
Clouds relative evolutionary changes of the radius do not exceed 10\%.

The general period--luminosity diagram involving all hydrodynamical models
computed in the present work is shown in Fig.~6.
In this diagram we represent 83 hydrodynamical models for $Z=0.008$
($6M_\odot\le\mzams\le 28M_\odot$)
and 19 models for $Z=0.004$ ($10M_\odot\le\mzams\le 24M_\odot$).
In Fig.~6 are plotted also 7 models for $Z=0.02$ with initial masses
$\mzams=15$ and $20M_\odot$.

The role of the mass loss rate $\dot M$ in the mass--luminosity relation
becomes perceptible only for stars with initial masses $\mzams\ge 20M_\odot$.
It is seen in Fig.~5 where two model sequences for $Z=0.008$ are shown
for the mass loss rate factor $f=0.3$ and $f=1$.
The mass loss rate increases with increasing luminosity (i.e. the initial mass)
of the red supergiant.
At the same time the role of nonlinear effects also increases with luminosity,
so that the scatter of values of the radial pulsation period $\Pi$ becomes
larger.
From hydrodynamic computations we found that the scatter due to nonlinear effects
is roughly comparable with effects of mass loss, that is why the models in Fig.~6
are shown for each value of $Z$ independently of the factor $f$.

Linear fits for the results of our hydrodynamic calculations in the
period--luminosity diagram are given by
\begin{equation}
\label{pl}
\begin{array}{ll}
\mbol = -0.575 - 2.832 \log \Pi , \quad & (Z=0.004) ,
\\[4pt]
\mbol = -0.337 - 2.820 \log \Pi ,       & (Z=0.008) ,
\end{array}
\end{equation}
where the period $\Pi$ is expressed in days.
For Magellanic Clouds ($0.004\le Z\le 0.008$) the dependence of the absolute
bolometric light $\mbol$ on $Z$ is approximately given by relation
\begin{equation}
\delta\mbol = 0.89 \delta\log Z ,
\end{equation}
which slightly differs from
$\delta\mbol = 0.83 \delta\log Z$
obtained by Guo and Li (2002).

\subsection*{period--mass diagram}

The most important application of the stellar pulsation theory is the
determination of the stellar mass with the period--mean density relation.
To this end together with the period of light variations $\Pi$
one should have the observational estimate of the mean radius
of the pulsating star $R$.
Unfortunately, for red supergiants such an approach is impossible
because of too large uncertainties in observational estimates of
their effective temperatures.
In particular, the problem is complicated due to both distorted continuum
because of numerous molecular absorption bands and circumstellar reddening
by dust grains condensing in the stellar wind.

However there is another approach to establish the relationship between
the stellar mass $M$ and the pulsation period $\Pi$ without observational
estimates of the stellar bolometric magnitude and effective temperature.
Such a relationship can be determined because the evolutionary changes of
the luminosity during thermonuclear helium burning proceed within
rather narrow ranges.
For example, for metal abundances $0.004\le Z\le 0.008$
the evolutionary luminosity variations range from
$\delta\log L = 0.08$ for $25M_\odot$ to $\delta\log L = 0.14$ for $\mzams=6$.

Evolution of the pulsationally unstable red supergiant in the
period--mass diagram is shown in Fig.~7 for two model sequences
with $\mzams=20M_\odot$ and metal abundances $Z=0.004$ and $Z=0.008$,
respectively.
In the general diagram shown in Fig.~8 all hydrodynamical models with
metal abundances $0.004\le Z\le 0.008$ are concentrated within the
strip with boundaries given by expressions
\begin{equation}
\label{pm}
\log (M/M_\odot) =
 \begin{Bmatrix}
  0.33 \\ 0.24
 \end{Bmatrix}
 + 0.368 \log\Pi ,
\end{equation}
where the period $\Pi$ is expressed in days.

\subsection*{conclusion}

Results of stellar evolution and nonlinear radial pulsation calculations show that
oscillations of red supergiants in the Magellanic Clouds are due to instability of
the fundamental mode.
In the HR diagram the pulsational instability domain encompasses the wide luminosity
range ($2.5\times 10^3 L_\odot\le L \le 2.7\times 10^5$) because both massive stars
($M > 10M_\odot$) and intermediate--mass stars ($6M_\odot\le M\le 10M_\odot$) are
pulsationally unstable.
The theoretical period--luminosity relation comprises more than one and a half orders
of magnitude ($17~\mbox{day}\le\Pi\le 1200~\mbox{day}$).
This result agrees well with classification of the General Catalogue of Variable Stars
(Samus et al. 2011) where the semiregular variables SR are mentioned as late type
giants and supergiants with light variation periods from 20 to 2000 days.
Observers studying the period--luminosity relation of red supergiants
in LMC and SMC dealt so far with stars with periods of light variation greater than 200 days,
whereas the results presented above allow us to conclude that the period--luminosity
relation extends to shorter periods.

Determination of masses of red supergiants with the period--mean density relation
is impossible because of large uncertainties in observational estimates of effective
temperatures.
However, due to enough small evolutionary changes of the stellar radius during the
thermonuclear helium burning the masses of red supergiants in LMC and SMC can be
evaluated using the observational estimate of the period of light variation
from the period--mass relation (\ref{pm}).
The uncertainty of this estimate $\delta\log M \approx 0.09$ (i.e. no more than 25\%)
is substantially less that for galactic red supergiants where the uncertainty
of mass determination is around 50\% (Fadeyev 2012).

The recent empirical period--luminosity relations for red supergiants in LMC and SMC
are determined in magnitudes of near--IR photometric bands and uncertain effective
temperatures of these stars do not allow the bolometric magnitudes to be correctly
determined.
Furthermore, there is considerable discrepancy between photometrical data obtained by
different observers (see, for example, Fig. 13 in the paper by Yang and Jiang (2011)).
That is why the theoretical period--luminosity relation in Fig.~6 was not compared with
observational data.

The study was supported by the Basic Research Program of the Russian Academy of Sciences
``Nonstationary phenomena in the Universe''.

\subsection*{REFERENCES}

\begin{enumerate}
\item E. B\"ohm--Vitense, Zeitschrift f\"ur Astrophys. \textbf{46}, 108 (1958).

\item Yu.A. Fadeyev, Pis'ma Astron. Zh. \textbf{33}, 775 (2007) [Astron. Lett. \textbf{33}, 692 (2007)].

\item Yu.A. Fadeyev, Pis'ma Astron. Zh. \textbf{36}, 380 (2010) [Astron. Lett. \textbf{36}, 362 (2010)].

\item Yu.A. Fadeyev, Pis'ma Astron. Zh. \textbf{37}, 440 (2011) [Astron. Lett. \textbf{37}, 403 (2011)].

\item Yu.A. Fadeyev, Pis'ma Astron. Zh. \textbf{38}, 295 (2012) [Astron. Lett. \textbf{38}, 260 (2012)].

\item I.S. Glass, MNRAS \textbf{186}, 317 (1979).

\item J.H. Guo and Y. Li, Astrophys.J. \textbf{565}, 559 (2002).

\item A. Heger, L. Jeannin, N. Langer, et al., Astron. Astrophys. \textbf{327}, 224 (1997).

\item C. de Jager, H. Nieuwenhuijzen, and K. A. van der Hucht, Astron. Astrophys. Suppl. Ser. \textbf{72}, 259 (1988).

\item J.S. Jurcevic, M.J. Pierce, and G.H. Jacoby, MNRAS \textbf{313}, 868 (2000).

\item R. Kuhfu\ss, Astron. Astrophys. \textbf{160}, 116 (1986).

\item Y. Li, Astron. Astrophys. \textbf{257}, 133 (1992a).

\item Y. Li, Astron. Astrophys. \textbf{257}, 145 (1992b).

\item Y. Li and Z.G. Gong, Astron. Astrophys. \textbf{289}, 449 (1994).

\item N. Mauron and E. Josselin, Astron. Astrophys. \textbf{526}, A156 (2011).

\item H. Nieuwenhuijzen and C. de Jager, Astron. Astrophys. \textbf{231}, 134 (1990).

\item M. Peimbert and S. Torres--Peimbert, Astrophys. J. \textbf{193}, 327 (1974).

\item M. Peimbert and S. Torres--Peimbert), Astrophys. J. \textbf{203}, 581 (1976).

\item M.J. Pierce, J.S. Jurcevic, and D. Crabtree, MNRAS \textbf{313}, 271 (2000).

\item S.B. Pope, \textit{Turbulent flows} (Cambridge University Press, 2000).

\item N.N. Samus, E.V. Kazarovets, N.N. Kireeva, et al. General Catalogue of Variable Stars (2011).

\item R. Stothers, Astrophys. J. \textbf{156}, 541 (1969).

\item R. Stothers, Astron. Astrophys. \textbf{18}, 325 (1972).

\item G. Wuchterl and M.U. Feuchtinger, Astron. Astrophys. \textbf{340}, 419 (1998).

\item M. Yang, B.W. Jiang, Astrophys. J. \textbf{727}, 53 (2011).

\item M. Yang, B.W. Jiang, Astrophys. J. \textbf{754}, 35 (2012).

\end{enumerate}

\newpage
\section*{FIGURE CAPTIONS}

\begin{itemize}
\item[Fig. 1.]
Evolutionary tracks of red supergiants $\mzams=20M_\odot$ with metal abundances
$Z=0.004$, 0.008 and 0.02 in the HR diagram.
Parts of tracks corresponding to thermonuclear core helium burning are shown in
solid lines, whereas parts of tracks with gravitational contraction as the
energy source in the stellar center are shown in dotted lines.
Arrows at the solid lines indicate the direction of evolution in the begining
of helium burning.
The mass loss factor $f$ is given at the upper ending points of the tracks.

\item[Fig. 2.]
The kinetic energy growth rate $\eta$ as a function of the central helium abundance
$\xhec$ in hydrodynamical models $\mzams=15$, 18, 20 and $25M_\odot$ with
$f=0.3$ and $Z=0.008$.
Hydrodynamical models are shown in filled circles.
Models of each evolutionary sequence are connected by dotted lines.

\item[Fig. 3.]
Variations of the gas flow velocity at the upper boundary $U$ (a)
and bolometric magnitude $\mbol$ (b) in the red supergiant with initial mass
$\mzams=20M_\odot$, central helium abundance $\xhec=0.39$ and
pulsation period $\Pi = 383$~days.

\item[Fig. 4.]
Same as Fig. 3 but for $\mzams=9M_\odot$, $\xhec=0.36$ and $\Pi = 52$~days.

\item[Fig. 5.]
Evolutionary changes of the luminoisity $L$ and the period of radial pulsations $\Pi$
in the red supergiants $\mzams=20M_\odot$ with metal abundances $Z=0.004$, 0.008 and 0.02.
Hydrodynamical models are represented by filled circles connected for each
evolutionary sequence by solid lines for $f=0.3$ and dotted lines for $f=1$.
Arrows indicate direction of evolutionary changes in the early stage of
thermonuclear helium burning.

\item[Fig. 6.]
The period--luminosity diagram for red supergiants.
Hydrodynamical models are represented by filled circles ($Z=0.008$), open circles
($Z=0.004$) and open triangles ($Z=0.02$).
Linear fits (\ref{pl}) are shown by dashed lines.

\item[Fig. 7.]
The period--mass diagram for red supergiants with initial mass $\mzams=20M_\odot$.
Hydrodynamical models are represented by filled circles connected by the dotted line
for each evolutionary sequence ($Z=0.004$ и $Z=0.008$).
The arrows indicate the direction of evolution in the diagram.

\item[Fig. 8.]
The period--mass diagram for red supergiants of Magellanic Clouds.
Hydrodynamical models with metal abandances $Z=0.008$ are $Z=0.004$
are shown by filled and open circles, respectively.
Boundaries of the red supergiant strip fitted by relations (\ref{pm})
are shown by dashed lines.

\end{itemize}

\newpage
\begin{figure}
\centerline{\includegraphics[width=15cm]{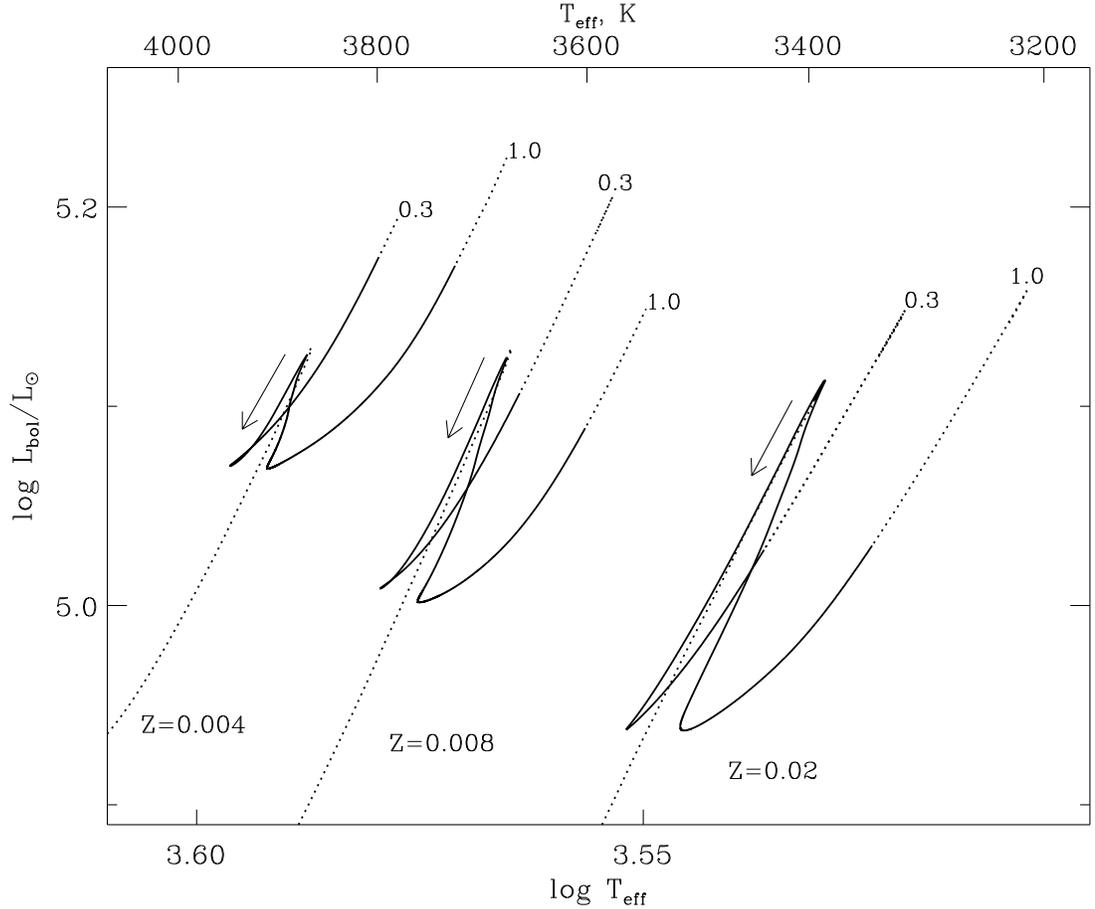}}
\caption{Evolutionary tracks of red supergiants $\mzams=20M_\odot$ with metal abundances
$Z=0.004$, 0.008 and 0.02 in the HR diagram.
Parts of tracks corresponding to thermonuclear core helium burning are shown in
solid lines, whereas parts of tracks with gravitational contraction as the
energy source in the stellar center are shown in dotted lines.
Arrows at the solid lines indicate the direction of evolution in the begining
of helium burning.
The mass loss factor $f$ is given at the upper ending points of the tracks.
}
\label{fig1}
\end{figure}

\newpage
\begin{figure}
\centerline{\includegraphics[width=15cm]{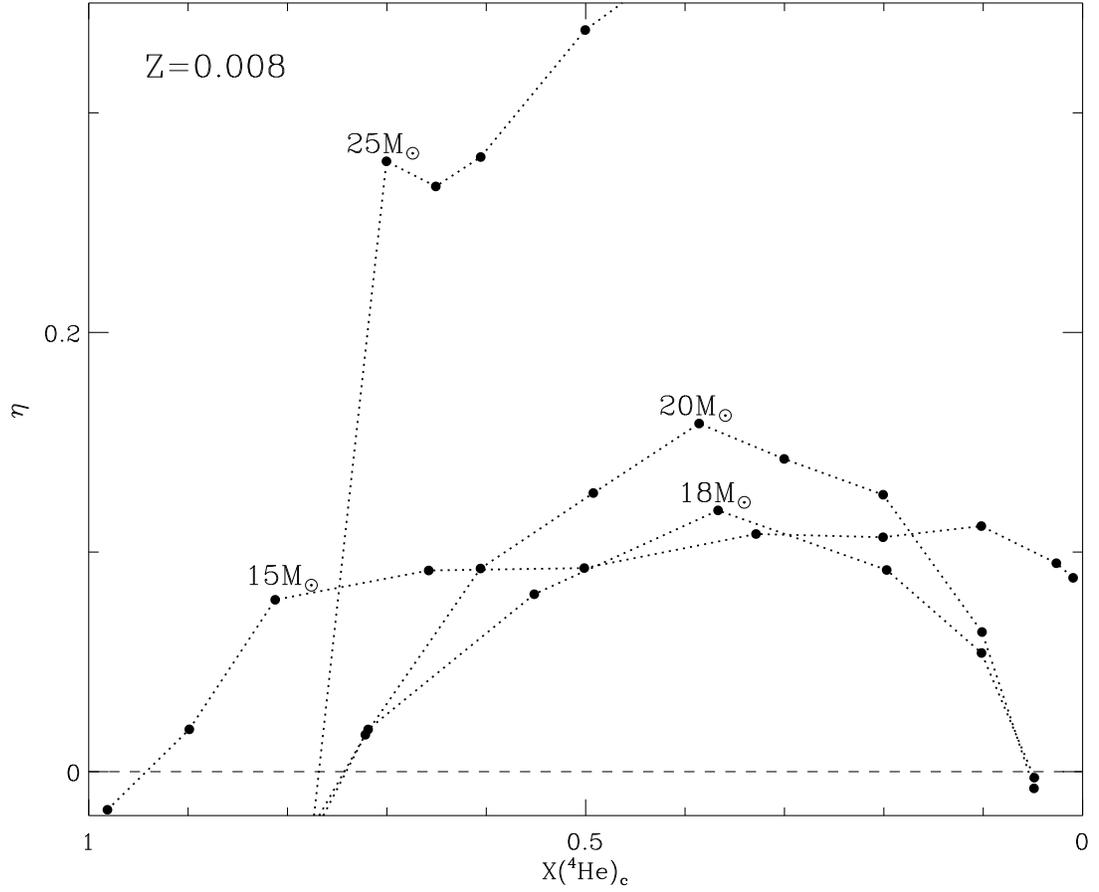}}
\caption{The kinetic energy growth rate $\eta$ as a function of the central helium abundance
$\xhec$ in hydrodynamical models $\mzams=15$, 18, 20 and $25M_\odot$ with
$f=0.3$ and $Z=0.008$.
Hydrodynamical models are shown in filled circles.
Models of each evolutionary sequence are connected by dotted lines.
}
\label{fig2}
\end{figure}

\newpage
\begin{figure}
\centerline{\includegraphics[width=15cm]{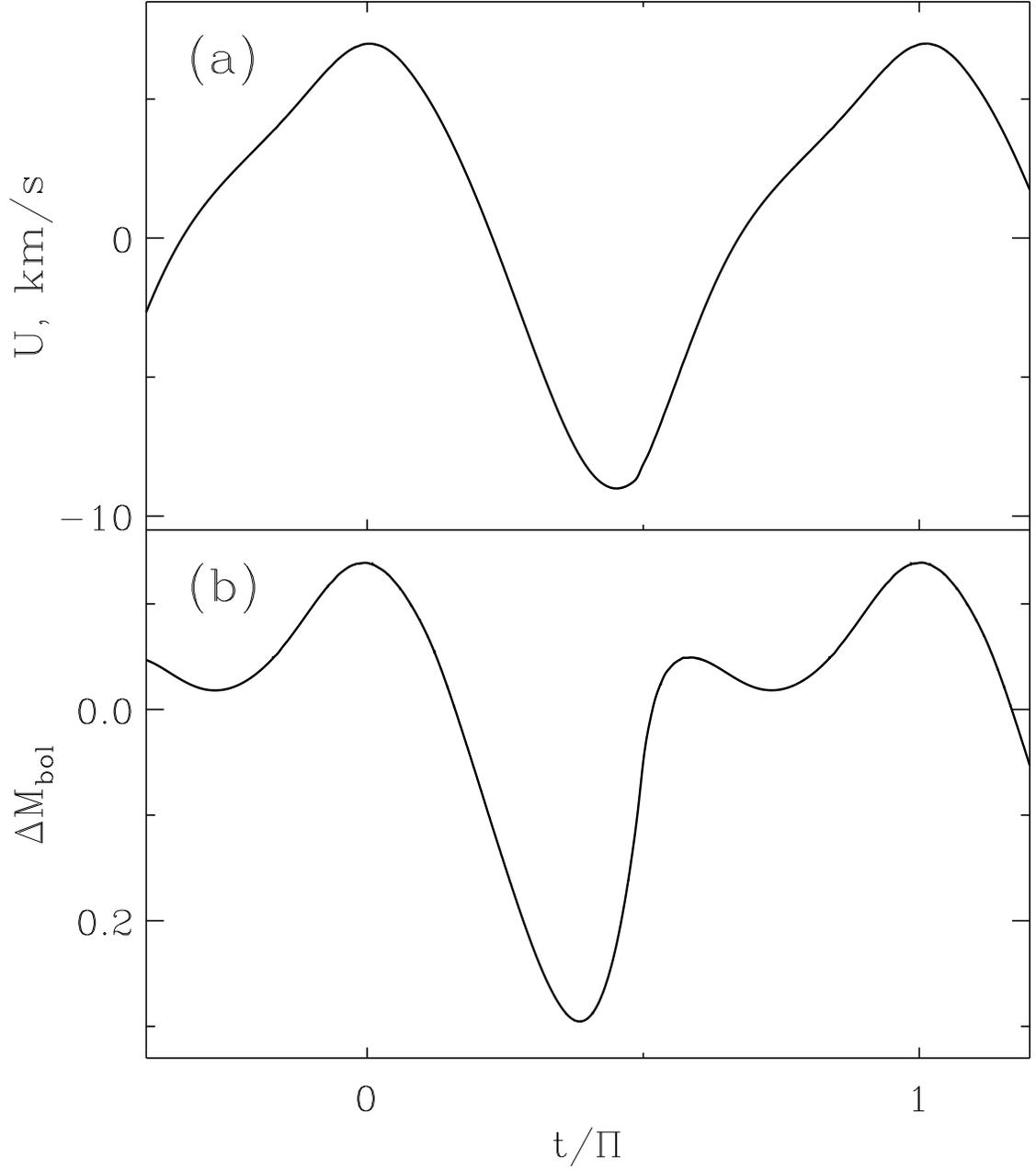}}
\caption{Variations of the gas flow velocity at the upper boundary $U$ (a)
and bolometric magnitude $\mbol$ (b) in the red supergiant with initial mass
$\mzams=20M_\odot$, central helium abundance $\xhec=0.39$ and
pulsation period $\Pi = 383$~days.
}
\label{fig3}
\end{figure}

\newpage
\begin{figure}
\centerline{\includegraphics[width=15cm]{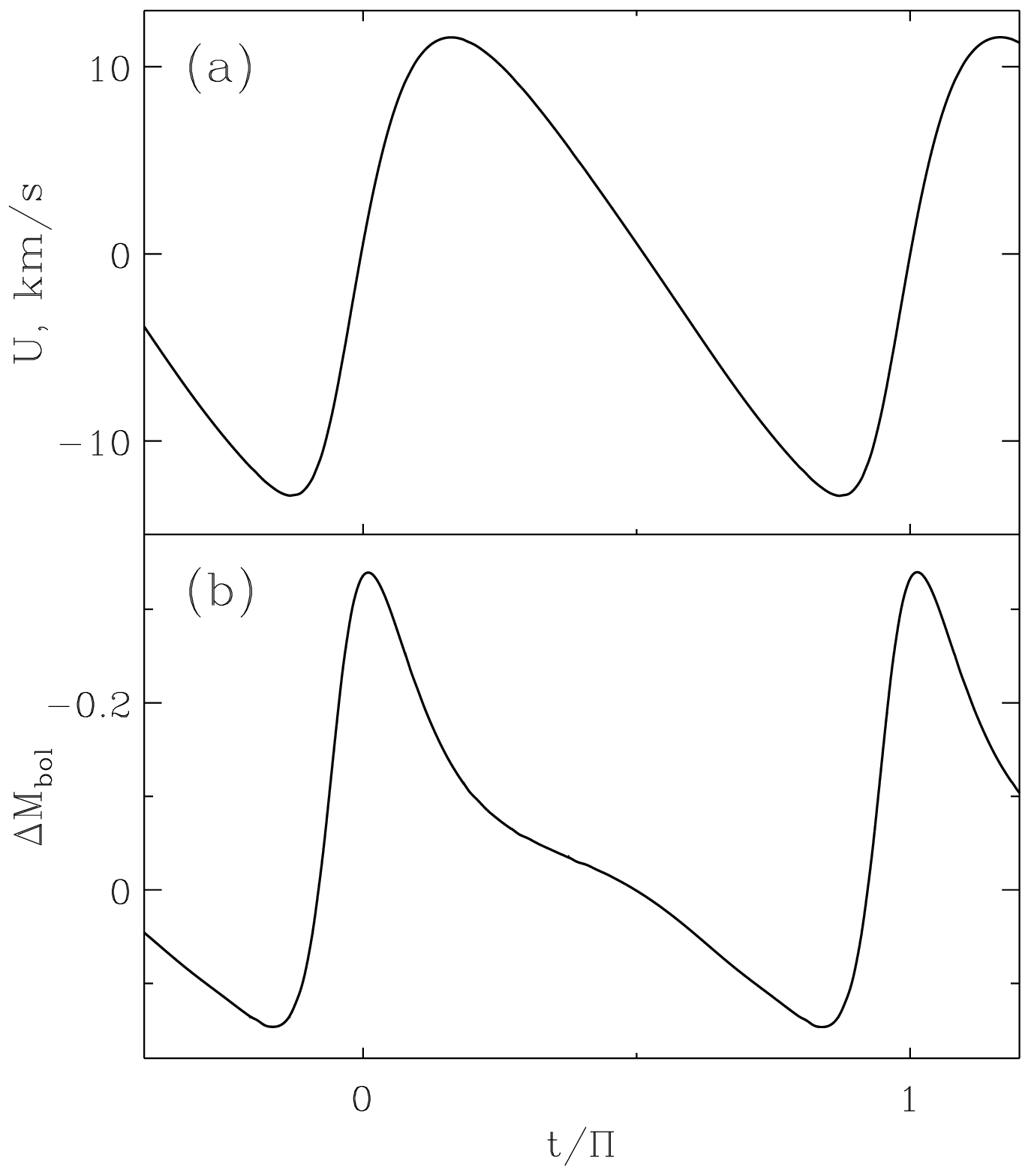}}
\caption{Same as Fig. 3 but for $\mzams=9M_\odot$, $\xhec=0.36$ and $\Pi = 52$~days.
}
\label{fig4}
\end{figure}

\newpage
\begin{figure}
\centerline{\includegraphics[width=15cm]{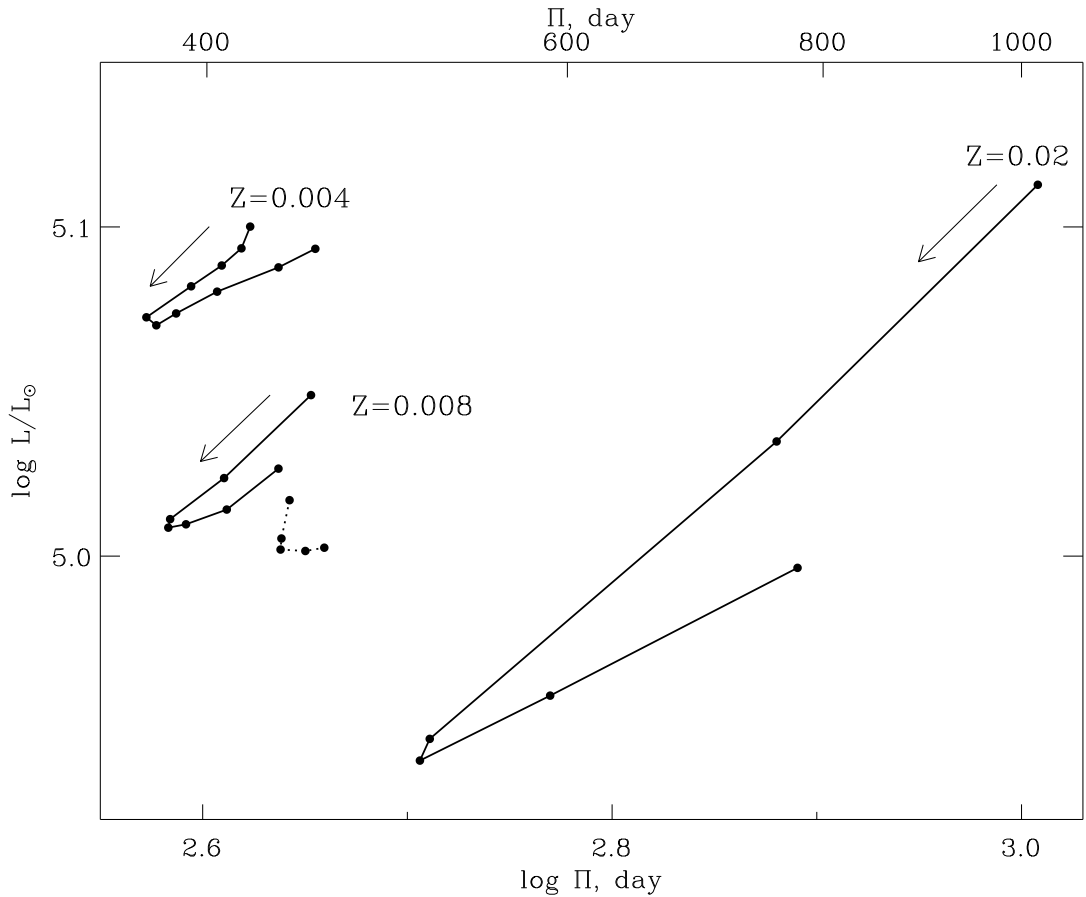}}
\caption{Evolutionary changes of the luminoisity $L$ and the period of radial pulsations $\Pi$
in the red supergiants $\mzams=20M_\odot$ with metal abundances $Z=0.004$, 0.008 and 0.02.
Hydrodynamical models are represented by filled circles connected for each
evolutionary sequence by solid lines for $f=0.3$ and dotted lines for $f=1$.
Arrows indicate direction of evolutionary changes in the early stage of
thermonuclear helium burning.
}
\label{fig5}
\end{figure}

\newpage
\begin{figure}
\centerline{\includegraphics[width=15cm]{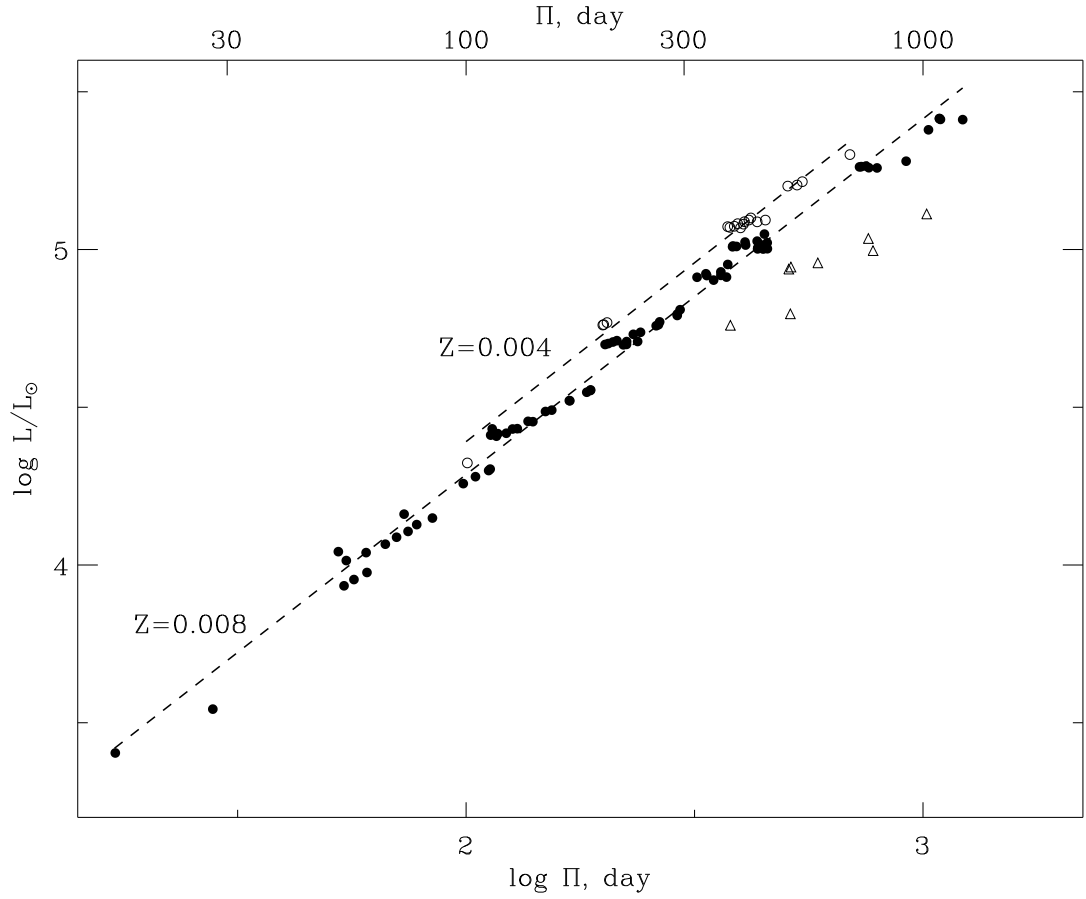}}
\caption{The period--luminosity diagram for red supergiants.
Hydrodynamical models are represented by filled circles ($Z=0.008$), open circles
($Z=0.004$) and open triangles ($Z=0.02$).
Linear fits (\ref{pl}) are shown by dashed lines.
}
\label{fig6}
\end{figure}

\newpage
\begin{figure}
\centerline{\includegraphics[width=15cm]{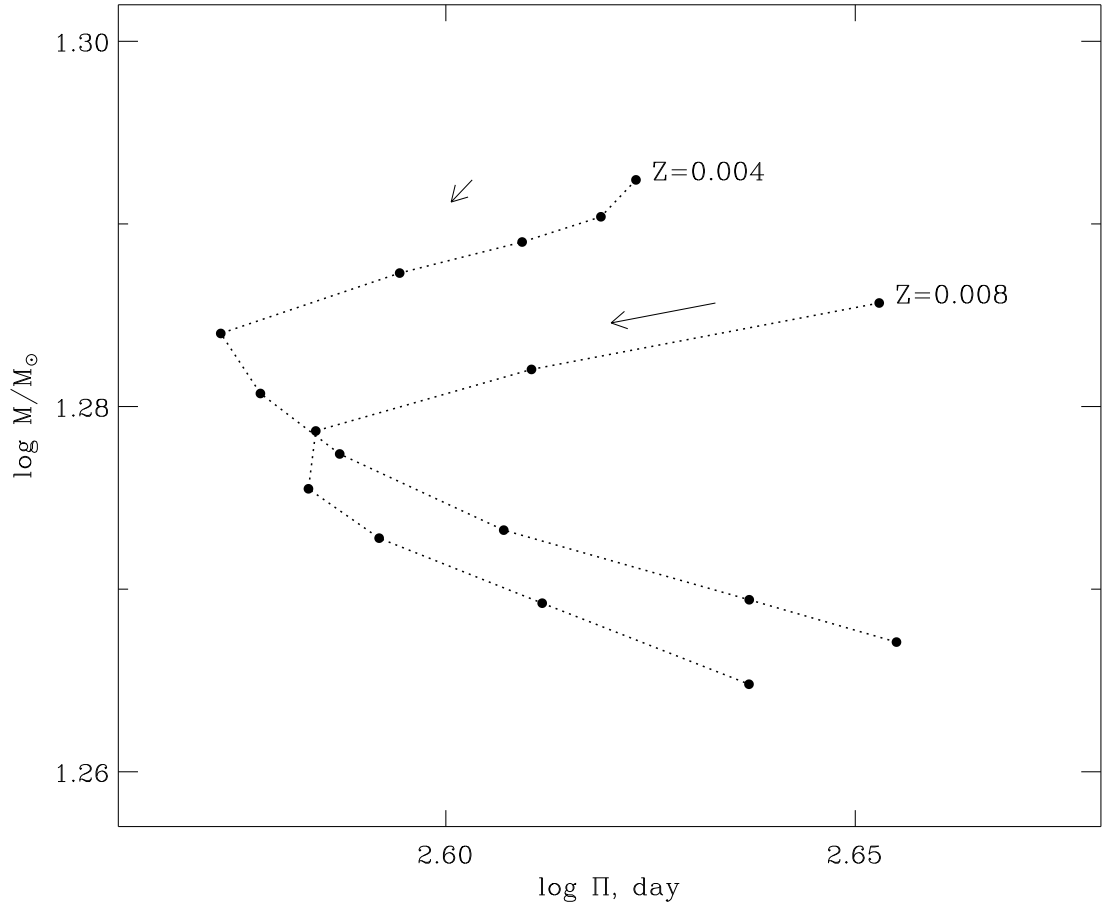}}
\caption{The period--mass diagram for red supergiants with initial mass $\mzams=20M_\odot$.
Hydrodynamical models are represented by filled circles connected by the dotted line
for each evolutionary sequence ($Z=0.004$ и $Z=0.008$).
The arrows indicate the direction of evolution in the diagram.
}
\label{fig7}
\end{figure}

\newpage
\begin{figure}
\centerline{\includegraphics[width=15cm]{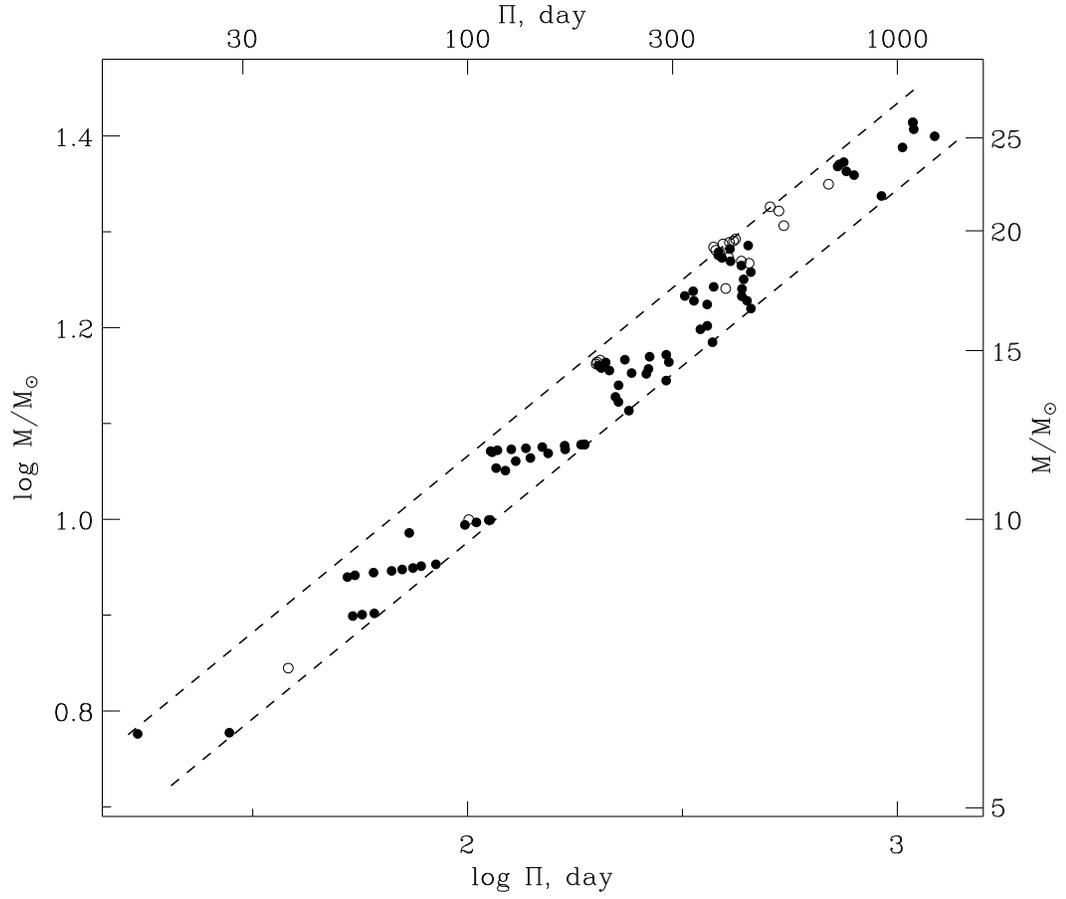}}
\caption{The period--mass diagram for red supergiants of Magellanic Clouds.
Hydrodynamical models with metal abandances $Z=0.008$ are $Z=0.004$
are shown by filled and open circles, respectively.
Boundaries of the red supergiant strip fitted by relations (\ref{pm})
are shown by dashed lines.
}
\label{fig8}
\end{figure}

\end{document}